\documentclass{article}
\usepackage{epsf}

\begin{document}
\title{Signals of the Abelian $Z'$ boson within the model
independent analysis of the LEP data}
\author{A. V. Gulov
 \thanks{Email: gulov@ff.dsu.dp.ua} and
 V. V. Skalozub
 \thanks{Email: skalozub@ff.dsu.dp.ua}\\
{\it Dniepropetrovsk National University,}\\ {\it Dniepropetrovsk,
49050 Ukraine}}
\date{April 4, 2002}
\maketitle
\begin{abstract}
The preliminary LEP data set on the total cross sections and the
forward-backward asymmetries of the $e^+ e^- \rightarrow \bar{f}f$
processes are analysed to establish a model-independent search for
the signals of the virtual Abelian $Z'$ boson. The recently
introduced observables giving possibility to pick up uniquely the
$Z'$ signals in these processes are used. The mean values of the
observables are in accordance with the $Z'$ detection at the
$1\sigma$ accuracy. The results of other model-independent fits
and further prospects are discussed.
\end{abstract}

\section{Introduction}

The recently finished LEP2 experiments have accumulated a huge
amount of data allowing to verify the predictions of the Standard
Model (SM) of elementary particles as well as to estimate the
energy scale of new physics beyond the SM. Although the complete
set of the LEP2 data is still being combined, the preliminary
total cross sections and the forward-backward asymmetries have
already been adduced in the literature \cite{EWWG}.

In the present note we are going to discuss the problem of
searching for signals of the heavy Abelian $Z'$ boson \cite{leike}
by means of analysis of the LEP data on the $e^+ e^- \rightarrow
\bar{f}f$ processes. This particle is the necessary element of
different models extending the SM. The low limits on its mass
estimated for the variety of popular models ($\chi$, $\psi$,
$\eta$, L--R models \cite{models} and the Sequential Standard
Model (SSM) \cite{SSM}) are found to be in the energy interval
500--2000 GeV \cite{EWWG} (see Table \ref{t1} which reproduces
Table 8.9 of Ref. \cite{EWWG}). As it is seen, the values of
$m_{Z'}$ (as well as the $Z'$ couplings to the SM particles) are
strongly model-dependent.

Other approach to find signals of new physics beyond the SM is the
model-independent analysis. In this consideration a number of
different contact interactions is introduced instead of specifying
the heavy particle content. Since only one parameter of new
physics can be successfully fitted, the LEP Collaborations usually
discuss eight `models' (LL, RR, LR, RL, VV, AA, A0, V0) which
assume specific helicity couplings between the initial state and
the final state fermion currents. Each model is described by only
one non-zero four-fermion coupling, while others are set to zero.
For example, in the LL model the non-zero coupling of left-handed
fermions is taken into account. The signal of a new heavy particle
is fitted by considering the interference of the SM amplitude with
the contact four-fermion term. Whatever physics beyond the SM
exists, it has to manifest itself in some contact coupling
mentioned. Hence, it is possible to find a low limit on the masses
of the states responsible for the interactions considered. In
principle, a number of states may contribute into each of the
models. Therefore, the purpose of the fit described by these
models is to find any signal of new physics. No specific types of
new particles are considered in these models. However, the virtual
states of a realistic heavy particle (for instance, the Abelian
$Z'$ boson) contribute to several contact interactions
simultaneously and the corresponding couplings cannot be switched
off separately.

Notice the fits for the process $e^+e^-\to\mu^+\mu^-$ in Ref.
\cite{EWWG}. A half of the eight models mentioned demonstrates the
one standard deviation from the SM predictions. In this regard, we
note Ref. \cite{bourilkov} where these models were applied to the
Bhabha scattering $e^+e^-\to e^+e^-(\gamma)$. The deviations from
the SM at the $1\sigma$--level were found again, whereas the AA
model shows the $2\sigma$--level deviation. However, these
deviations could not be interpreted as the signal of the Abelian
$Z'$ boson.

Because of the mentioned arguments it seems to us reasonable to
find some model independent signals of the $Z'$ boson. To
elaborate that we have taken into consideration some general
principles of the field theory which give possibility to relate
the parameters of different scattering processes. Then we
introduced the variables, convenient to pick up uniquely the $Z'$
boson (or other heavy states). In Ref. \cite{ZprTHDM} the
model-independent sign-definite observables for the Abelian $Z'$
detection in four-fermion scattering processes at $\sqrt{s}\simeq
500$ GeV were introduced.

As it was pointed out in Ref. \cite{ZprTHDM}, some $Z'$ couplings
to the SM particles could be related by using the requirement of
renormalizability of the underlying model remaining in other
respects unspecified. The relations between the parameters of new
physics appearing due to the renormalizability were called the
renormalization group (RG) relations. The derived in Ref.
\cite{ZprTHDM} RG relations predict two possible types of the
low-energy $Z'$ interactions with the SM fields, namely, the
chiral and the Abelian $Z'$ bosons. Each $Z'$ type is described by
a few couplings to the SM fields. Therefore, it is possible to
introduce observables which uniquely pick up the $Z'$ virtual
state \cite{ZprTHDM}. In the present paper we discuss the
observables at LEP energies and the constraints on possible
signals of Abelian $Z'$-boson following from the analysis of the
LEP data. The content of the paper is the following. In sect. 2
the necessary information on the model-independent description of
the $Z'$ interactions at low energies and the RG relations are
given. In sect. 3 the observables to pick up uniquely the $Z'$
boson are introduced. In the last section the results on the LEP
data fit and the conclusions as well as further prospects are
discussed.

\section{$Z'$ couplings to fermion and scalar fields}

The Abelian $Z'$ boson can be introduced in a phenomenological way
by defining its effective low-energy couplings to the SM fields.
Such a parameterization is well known in the literature
\cite{leike}. Since we are going to account of the $Z'$ effects in
the low-energy $e^+e^-\to l^+l^-$ processes, we consider the
tree-level $Z'$ interactions, only. As the decoupling theorem
\cite{decoupling} guarantees, they are of renormalizable type,
since the non-renormalizable interactions are generated at higher
energies due to radiation corrections and suppressed by the
inverse heavy mass. The SM gauge group $SU(2)_L\times U(1)_Y$ is
considered as a subgroup of the underlying theory group. So, the
mixing interactions of the types $Z'W^+W^-$, $Z'ZZ$, ... are
absent at the tree level. Under these assumptions the $Z'$
couplings to the fermion and scalar fields are described by the
Lagrangian:
\begin{eqnarray}\label{1}
 {\cal L}&=& \left|\left( D^{{\rm ew,} \phi}_\mu -
  \frac{i\tilde{g}}{2}\tilde{Y}(\phi)\tilde{B}_\mu
  \right)\phi\right|^2 + \nonumber\\&&
 i\sum\limits_{f=f_L,f_R}\bar{f}{\gamma^\mu}
  \left(
  D^{{\rm ew,} f}_\mu -
  \frac{i\tilde{g}}{2}\tilde{Y}(f)\tilde{B}_\mu
  \right)f,
\end{eqnarray}
where $\phi$ is the SM scalar doublet, $\tilde{B}_\mu$ denotes the
massive $Z'$ field before the spontaneous breaking of the
electroweak symmetry, and the summation over the all SM
left-handed fermion doublets, $f_L =\{(f_u)_L, (f_d)_L\}$, and the
right-handed singlets, $f_R = (f_u)_R, (f_d)_R$, is understood.
The notation $\tilde{g}$ stands for the charge corresponding to
the $Z'$ gauge group, $D^{{\rm ew,}\phi}_\mu$ and $D^{{\rm
ew,}f}_\mu$ are the electroweak covariant derivatives. Diagonal
$2\times 2$ matrices $\tilde{Y}(\phi)={\rm
diag}(\tilde{Y}_{\phi,1},\tilde{Y}_{\phi,2})$,
$\tilde{Y}(f_L)={\rm diag}(\tilde{Y}_{L,f_u},\tilde{Y}_{L,f_d})$
and numbers $\tilde{Y}(f_R)=\tilde{Y}_{R,f}$ are unknown $Z'$
generators characterizing the model beyond the SM.

In particular, the Lagrangian (\ref{1}) describes the $Z$--$Z'$
mixing of order $m^2_Z/m^2_{Z'}$ which is proportional to
$\tilde{Y}_{\phi,2}$ and originated by the diagonalization of the
neutral boson states. The mixing contributes to the scattering
amplitudes and cannot be neglected at the LEP energies \cite{NPB}.

Thus, the $Z'$ couplings to any fermion $f$ are parameterized by
the numbers $\tilde{Y}_{L,f}$ and $\tilde{Y}_{R,f}$.
Alternatively, one can use the couplings to the axial-vector and
vector fermion currents,
$a^l_{Z'}\equiv(\tilde{Y}_{R,l}-\tilde{Y}_{L,l})/2$ and
$v^l_{Z'}\equiv(\tilde{Y}_{L,l}+\tilde{Y}_{R,l})/2$.

The parameters $a_f$, $v_f$ are usually treated as independent
numbers. However, they are related if the underlying theory is
renormalizable. The detailed discussion of this point as well as
the derivation of the RG relations are given in Ref.
\cite{ZprTHDM}. Therein it is shown that two possible types of the
$Z'$ bosons are possible -- the chiral and the Abelian ones. In
the present paper we are interested in the Abelian $Z'$ couplings
which are described by the relations:
\begin{equation}\label{abelian}
v_f-a_f=v_{f^\star}-a_{f^\star},\quad a_f=T_{3,f}\tilde{Y}_\phi,
\quad \tilde{Y}_{\phi,1}=\tilde{Y}_{\phi,2}\equiv\tilde{Y}_\phi
\end{equation}
where $T^3_f$ is the third component of the fermion weak isospin,
and $f^\star$ means the isopartner of $f$ (namely,
$l^\star=\nu_l,\nu^\star_l=l,\ldots$). The relations
(\ref{abelian}) ensure, in particular, the invariance of the
Yukawa terms with respect to the effective low-energy
$\tilde{U}(1)$ subgroup corresponding to the Abelian $Z'$ boson.
As it follows from the relations, the couplings of the Abelian
$Z'$ to the axial-vector fermion currents have the universal
absolute value proportional to the $Z'$ coupling to the scalar
doublet. So, in what follows we will use the short notation
$a=a_l=-\tilde{Y}_\phi/2$.

Notice that the same relations (\ref{abelian}) hold in the
two-Higgs-doublet model (THDM) \cite{THDM}. As a consequence, the
results of the present note are also valid for the case of the
THDM as the low-energy theory.

Because of a fewer number of independent $Z'$ couplings it is
possible to introduce the observables convenient for detecting
uniquely the $Z'$ signals in experiments. In what follows, we take
into account the RG relations (\ref{abelian}) in order to
constrain signals of the Abelian $Z'$ boson.

\section{Observables}

Consider the lepton processes $e^+e^-\to V^\ast\to l^+l^-$
($l=\mu,\tau$) with the neutral vector boson exchange
($V=A,Z,Z'$). We assume the non-polarized initial- and final-state
fermions. At LEP energies $\sqrt{s}\simeq 200$ GeV the leptons can
be treated as massless particles. In this approximation the
left-handed and the right-handed fermions can be substituted by
the helicity states.

To take into consideration the correlations (\ref{abelian}) let us
introduce the observable $\sigma_l(z)$ defined as the difference
of cross sections integrated in some ranges of the scattering
angle $\theta$:
\begin{eqnarray}\label{eq8}
 \sigma_l(z)
 &\equiv&\int\nolimits_z^1
  \frac{d\sigma_l}{d\cos\theta}d\cos\theta
 -\int\nolimits_{-1}^z
  \frac{d\sigma_l}{d\cos\theta}d\cos\theta
 \nonumber\\&=&
 \sigma^T_l\left[ A^{FB}_l\left(1-z^2\right)
 -\frac{z}{4}\left(3+z^2\right)\right],
 \nonumber\\
\Delta\sigma_l (z) &\equiv& \sigma_l (z) -\sigma^{\rm SM}_l (z),
\end{eqnarray}
where $z$ stands for the cosine of the boundary angle,
$\sigma^T_l$ denotes the total cross section and $A^{FB}_l$ is the
forward-backward asymmetry of the process. The idea of introducing
the $z$-dependent observable (\ref{eq8}) is to choose the value of
the kinematic parameter $z$ in such a way that to pick up the
characteristic features of the Abelian $Z'$ signals. Since the
observable $\Delta\sigma_l (z)$ is a small quantity, it can be
computed in lower order by the Born amplitudes for $e^+e^-\to
V^\ast\to l^+l^-$ $(V=\gamma,Z,Z')$.

The expansion of the $Z'$-boson propagator $(s-m^2_{Z'})^{-1}$ and
the $Z$--$Z'$ mixing angle in the inverse $Z'$ mass produces a
number of terms of order $m^{-2}_{Z'}$ and higher. The lower-level
contributions describe the four-fermion contact interactions and
contain the ratio $\tilde{g}^2/m^2_{Z'}$ of the $Z'$ mass and the
charge $\tilde{g}$, only. Thus, the quantities $m_{Z'}$ and
$\tilde{g}$ cannot be measured separately by the fit of
observables in the leading order in $m^{-2}_{Z'}$. In what follows
we will also treat the terms of order $m^{-4}_{Z'}$. As we will
show, these contributions allow one to fit both the four-fermion
coupling constant $\tilde{g}^2/m^2_{Z'}$ and the $Z'$ mass, if the
cross sections at different center-of-mass energies are taken into
account.

Due to the correlations between the Abelian $Z'$ couplings the
cross section (\ref{eq8}) can be written as follows
\begin{equation}\label{obs2}
 \Delta\sigma_l(z)=\sum_{j=0}^2
 \left[A_j(s,z)+\zeta B_j(s,z) \right]\epsilon_j +
 \sum_{j=0}^2\sum_{k=0}^j C_{jk}(s,z)\epsilon_j \epsilon_k,
\end{equation}
where we introduce the dimensionless quantities
\begin{eqnarray}\label{5}
 \epsilon_0 &=& \frac{\tilde{g}^2 m^2_Z a^2}{4\pi m^2_{Z'}},\quad
 \epsilon_1 = \frac{\tilde{g}^2 m^2_Z v_e v_l}{4\pi m^2_{Z'}},
\nonumber\\
 \epsilon_2 &=& \frac{\tilde{g}^2 m^2_Z a (v_e+v_l)}{4\pi m^2_{Z'}},
 \quad\zeta = \frac{m^2_Z}{m^2_{Z'}}.
\end{eqnarray}
The functions $A_j(z,s)$, $B_j(z,s)$ and $C_{jk}(z,s)$ are
determined by the SM couplings and masses, only. They are also
independent of the lepton generation. The factors $A_j$ describe
the leading-order contributions, whereas others correspond to the
higher-order corrections in the inverse $Z'$ mass.

As it was argued in Refs. \cite{NPB,obs}, there is a region of
values $z$, at which all the factors $A_j$ except for $A_0$
contribute less than 2\%. Since the parameter $\epsilon_0$ is a
positive quantity by the definition, it is possible to construct a
sign-definite observable by specifying the appropriate value of
the kinematic parameter $z$. This value, $z=z^*$, can be chosen in
order to maximize the relative contribution of the sign-definite
terms in $\Delta\sigma_l(z)$. To take into account the order of
each term in the inverse $Z'$ mass, we introduce positive
`weights' $\omega_B\sim\zeta$ and $\omega_C\sim |\epsilon_j|$ for
the higher-order contributions. Thus, $z^*$ is found by the
maximization of the following function:
\begin{equation}
F=\frac{|A_0|+\omega_B |B_0| + \omega_C |C_{00}|}
{\sum_{j=0}^2 \left(|A_j|+\omega_B |B_j| +
\sum_{k=0}^j\omega_C |C_{jk}|\right)}.
\end{equation}
The numeric values of the `weights' $\omega_B$ and $\omega_C$ can
be taken from the present day bounds on the contact couplings
\cite{EWWG} or \cite{hep01}. As the computation shows, the value
of $z^*$ with the accuracy $10^{-3}$ depends on the order of the
`weight' magnitudes, only. So, in what follows we take
$\omega_B\sim .004$ and $\omega_C\sim 0.00004$.

The function $z^\ast(s)$ is the decreasing function of the
center-of-mass energy. It is tabulated for the LEP energies in
Table 2. The corresponding values of the maximized function $F$
are within the interval $0.960<F<0.966$.

Since $A_0(s,z^*)<0$, $B_0(s,z^*)<0$ and $C_{00}(s,z^*)<0$,
the observable
\begin{equation}\label{7}
 \Delta\sigma_l(z^*)=
 \left[A_0(s,z^*)+\zeta B_0(s,z^*) \right]\epsilon_0 +
 C_{00}(s,z^*)\epsilon^2_0
\end{equation}
is negative and the same for the all types of the SM charged
leptons with the accuracy 2--4\%. This observable selects the
model-independent signal of the Abelian $Z'$ boson in the
processes $e^+e^-\to l^+l^-$.

\section{Data fit and Conclusions}

To search for the model-independent $Z'$ signals we will analyze
the introduced observable $\Delta\sigma_l (z^\ast)$ on the base of
the LEP data set. In the lower order in $m^{-2}_{Z'}$ the
observable (\ref{7}) depends on the one parameter, $\epsilon_0$,
\begin{equation}
 \Delta\sigma^{\rm th}_l(z^*)=
 A_0(s,z^*)\epsilon_0 + C_{00}(s,z^*)\epsilon^2_0,
\end{equation}
which can be fitted from the experimental values of
$\Delta\sigma_l (z^\ast)$. This approach has the following
advantages:
\begin{enumerate}
\item
All the LEP data for the lepton processes $e^+e^-\to l^+l^-$ can
be incorporated to obtain the limits on the same
flavour-independent scale.
\item
The sign of the fitted parameter ($\epsilon_0 >0$) is the
characteristic feature of the Abelian $Z'$ signal.
\end{enumerate}

\begin{figure}\centering
\epsfxsize=0.4\textwidth \epsfbox[0 0 500 600]{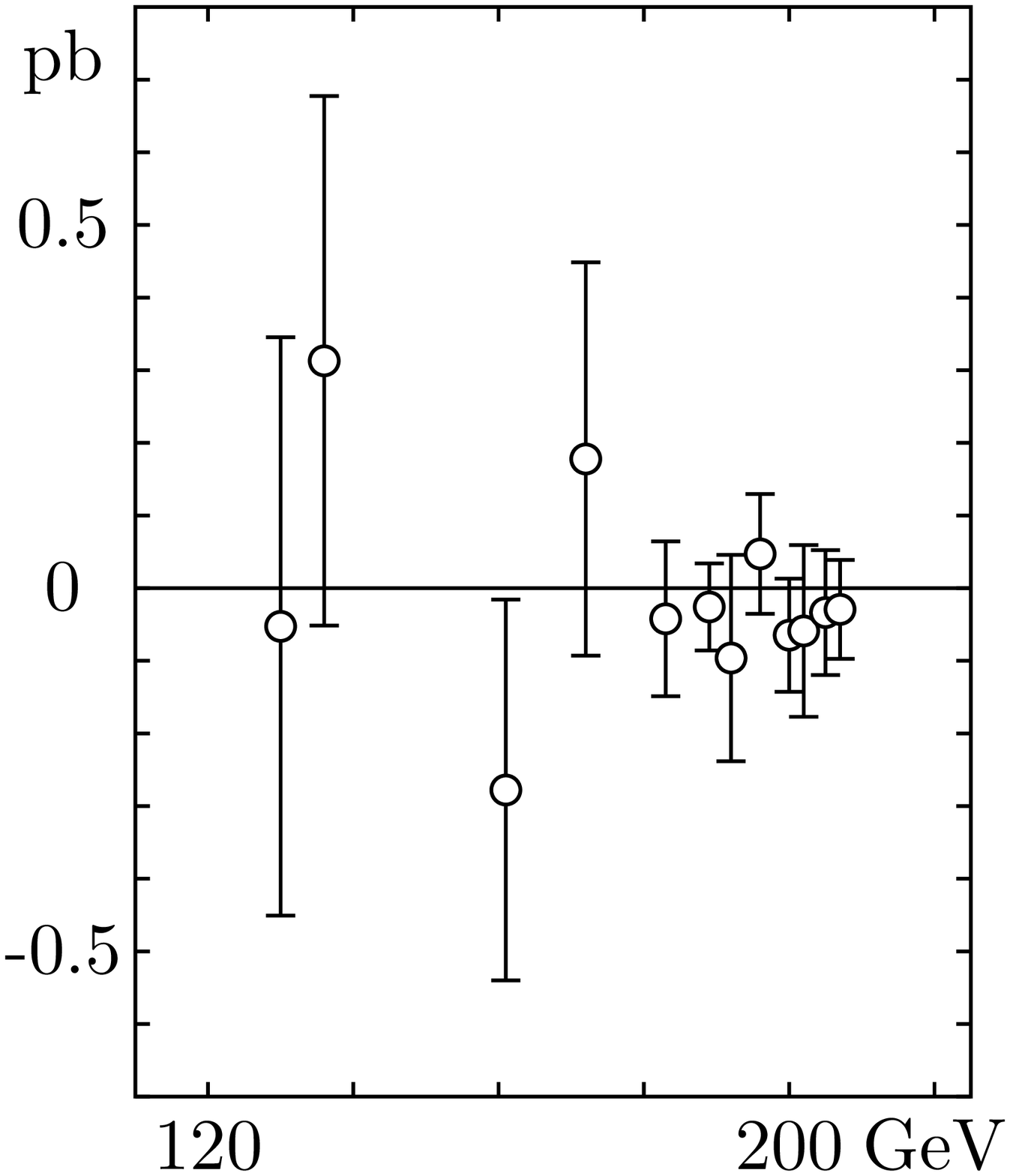}
\caption{$\Delta\sigma_\mu(z^\ast)$ computed from the LEP data.}
\label{fig:mu}
\end{figure}

\begin{figure}\centering
\epsfxsize=0.4\textwidth \epsfbox[0 0 400 600]{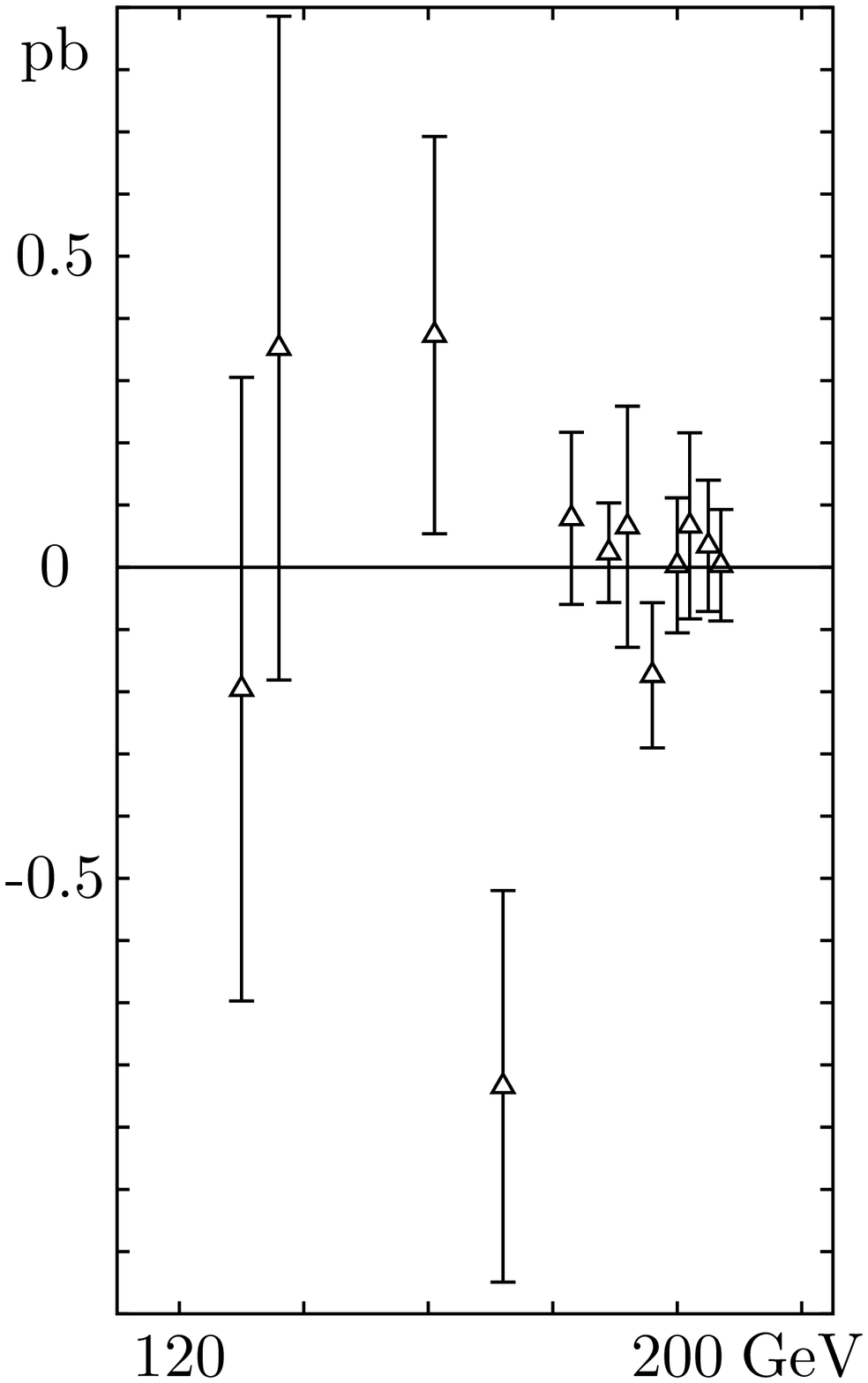}
\caption{$\Delta\sigma_\tau(z^\ast)$ computed from the LEP data.}
\label{fig:tau}
\end{figure}

The LEP data for the total cross-sections and the forward-backward
asymmetries \cite{EWWG} are converted to the experimental values
of the observable $\Delta\sigma_l(z^\ast)$ with the corresponding
errors $\delta\sigma_l(z^\ast)$ for each LEP energy by means of
the following relations:
\begin{eqnarray}%\label{}
 \Delta\sigma_l(z^\ast)
 &=&
 \left[
 A_l^{FB}\left(1-z^{\ast 2}\right)
 -\frac{z^\ast}{4}\left(3 +z^{\ast 2}\right)
 \right] \Delta\sigma_l^T
 \nonumber\\&&
 + \left(1 - z^{\ast 2}\right)
  \sigma_{l,\rm SM}^T \Delta A_l^{FB},
 \nonumber\\
 \delta\sigma_l(z^\ast)^2
 &=&
 {\left[
 A_l^{FB}\left(1-z^{\ast 2}\right)
 -\frac{z^\ast}{4}\left(3 +z^{\ast 2}\right)
 \right]}^2 (\delta\sigma_l^T)^2
 \nonumber\\&&
 +{\left[
 \left(1 - z^{\ast 2}\right)
 \sigma_{l,\rm SM}^T
 \right]}^2 (\delta A_l^{FB})^2.
\end{eqnarray}
The results are given in Table 2 and Figs.
\ref{fig:mu}--\ref{fig:tau}.

As it is seen, all the values of the observable are no more than
one standard deviation from the SM value $\Delta\sigma_l(z^*)=0$
except for the value of $\Delta\sigma_\mu$ at 161 GeV and three
points at 161, 172 and 196 GeV corresponding to the
$e^+e^-\to\tau^+\tau^-$ process. These points reflect the
significant dispersion of the measurements at $\sqrt{s}< 183$ GeV.
As it is also seen from Figs. \ref{fig:mu}--\ref{fig:tau}, the
measurements for the scattering into $\mu$ pairs have a higher
level of precision. Thus, in what follows we will use two sets of
data: 12 points for the $e^+e^-\to\mu^+\mu^-$ process and the full
data set including 24 points for $\mu^+\mu^-$ and $\tau^+\tau^-$
in the final state.

The central value of the fitted parameter $\epsilon_0$ is obtained
as the result of minimization of the $\chi^2$-function:
\begin{equation}
\chi^2(\epsilon_0) = \sum_{n} \frac{\left[\Delta\sigma^{\rm
ex}_{l,n}(z^*)- \Delta\sigma^{\rm th}_l(z^*)\right]^2}
{\delta\sigma^{\rm ex}_{l,n}(z^*)^2},
\end{equation}
where the sum runs over the experimental points entering the data
set chosen.

The $1\sigma$ confidence level interval $(b_1,b_2)$ for the fitted
parameter is derived by means of the likelihood function ${\cal
L}(\epsilon_0)\propto\exp[-\chi^2(\epsilon_0)/2]$. It is
determined by the equations:
\begin{equation}
\int\nolimits_{b_1}^{b_2}{\cal L}(\epsilon ')d\epsilon ' = 0.68,
 \quad
{\cal L}(b_1)={\cal L}(b_2).
\end{equation}

To compare our results with those of Ref. \cite{EWWG} we introduce
the contact interaction scale
\begin{equation}
\Lambda^2 = 4m^2_Z\epsilon^{-1}_0.
\end{equation}
This normalization of contact couplings is admitted in Ref.
\cite{EWWG}. We use the log-likelihood method to determine a one
sided lower limit on the scale $\Lambda$ at the 95\% confidence
level. It is derived by the integration of the likelihood function
over the physically allowed region $\epsilon_0>0$. The exact
definition is
\begin{equation}
\Lambda=2m_Z (\epsilon^*)^{-1/2}, \quad \int_{0}^{\epsilon^*}{\cal
L}(\epsilon ')d\epsilon ' = 0.95\int_{0}^{\infty}{\cal L}(\epsilon
')d\epsilon '.
\end{equation}

We also introduce the probability of the Abelian $Z'$ signal as
the integral of the likelihood function over the positive values
of $\epsilon_0$:
\begin{equation}
P=\int\nolimits_{0}^{\infty} L(\epsilon ')d\epsilon '.
\end{equation}

As it was mentioned above, we choose two different sets of data to
fit the parameter $\epsilon_0$. The first one includes
$e^-e^+\to\mu^-\mu^+$ scattering data (12 points), whereas the
second set includes both $e^-e^+\to\mu^-\mu^+$ and
$e^-e^+\to\tau^-\tau^+$ data (24 measurements). In Table \ref{t6}
we show the fitted values of $\epsilon_0$ with their 68\%
confidence level uncertainties, the 95\% confidence level lower
limit on the scale $\Lambda$, and the total probability of the
Abelian $Z'$ signal.

As it is seen, all the data sets lead to the comparable fitted
values of $\epsilon_0$ with the nearly equal uncertainties. All
the central values, $\bar\epsilon_0$, have the sign compatible
with the Abelian $Z'$ signal. The more precise data corresponding
to the scattering into $\mu^+\mu^-$ pairs demonstrate the largest
positive mean value of $\epsilon_0$. This value is at one standard
deviation from the SM prediction $\bar\epsilon_0=0$.

Taking into account the data for $\tau^+\tau^-$ final states
decreases the central value of $\epsilon_0$ but does not affect
essentially the uncertainty of the result. The corresponding
fitted value is no more than one standard deviation.

Thus, the fitted central values $\bar\epsilon_0$ witness to the
Abelian $Z'$ existence. The signal is at one standard deviation
for the $e^-e^+\to\mu^-\mu^+$ data. No signal is found at the
$1\sigma$ confidence level for the full lepton data set.

In fact, the fitted value of the contact coupling $\epsilon_0$
originates from the leading-order term in the inverse $Z'$ mass
contributing to the observable (\ref{7}). The analysis of the
higher-order terms allows to estimate the constraints on the $Z'$
mass alone. Substituting $\epsilon_0$ in Eq. (\ref{7}) by its
fitted central value from Table \ref{t6}, $\bar{\epsilon}_0$, one
obtains the expression
\begin{equation}
 \Delta\sigma_l(z^*)=
 \left[A_0(s,z^*)+\zeta B_0(s,z^*) \right]\bar\epsilon_0 +
 C_{00}(s,z^*)\bar\epsilon^2_0,
\end{equation}
which depends on the parameter $\zeta=m^2_Z/m^2_{Z'}$.

The central value of $\zeta$ and the $1\sigma$ confidence interval
can be computed in a way as those for $\epsilon_0$. The results
are also given in Table \ref{t6}.

Being governed by the next-to-leading contributions in
$m^{-2}_{Z'}$, the fitted values of $\zeta$ are characterized by
significant errors. The $\mu\mu$ data set gives the central value
which corresponds to $m_{Z'}\simeq 1.13$ TeV, whereas the full
lepton data set leads to the unphysical central value of $\zeta$.
Of course, the derived constraints on $\zeta$ are rather an
illustration of the possibility to fit the $Z'$ mass alone because
the higher-order terms in $m^{-2}_{Z'}$ have to be accounted for
simultaneously with the loop corrections to the factors $A_0$,
$B_0$ and $C_{00}$ in Eq. (\ref{7}). So, the analysis of the terms
$\sim m^{-4}_{Z'}$ requires to compute $\Delta\sigma_l(z^*)$ in
the improved Born approximation, which is the subject for separate
investigations. The important possibility to improve accuracy is
to use the data on the differential cross sections, when the
combined data on them will be completed. With these data taken
into consideration the observable $\Delta\sigma_l(z^*)$ can be
calculated directly from the definition Eq. (\ref{eq8}). In this
case the uncertainties have to decrease. We believe that all these
stages of the improvement of the data treating will make the
situation with the $Z'$ signals more transparent.

As it was shown, the characteristic signal of the Abelian $Z'$
boson is concerned with the coupling to axial-vector currents. In
this regard, let us turn again to the helicity `models' of Ref.
\cite{EWWG} and compare our results with the fit for the AA case.
As it follows from the present analysis, this model is sensitive
mainly to the signals of the Abelian $Z'$ boson. Of course, the
parameters $\epsilon$ in Ref. \cite{EWWG} and $\epsilon_0$ in Eq.
(\ref{5}) are not the same quantity. First, they are normalized by
different factors and related as $\epsilon=-\epsilon_0
m^{-2}_Z/4$. Second, as we already noted, in the AA model the $Z'$
couplings to the vector fermion currents are set to zero,
therefore it is able to describe only some particular case of the
Abelian $Z'$ boson. Moreover, in this model both the positive and
the negative values of $\epsilon$ are considered, whereas in our
approach only the positive $\epsilon_0$ values (which correspond
to the negative $\epsilon$) are permissible. As the value of the
four-fermion contact coupling in the AA model is dependent on the
lepton flavor, the Abelian $Z'$ induces the axial-vector coupling
which is universal for all lepton types. Nevertheless, it is
interesting to note that the fitted value of $\epsilon$ in the AA
model for the $\mu^+\mu^-$ final states
($-0.0025^{+0.0018}_{-0.0023}\mbox{ TeV}^{-2}$) as well as the
value derived under the assumption of the lepton universality
($-0.0018^{+0.0016}_{-0.0019}\mbox{ TeV}^{-2}$) are similar to our
results which correspond to $\epsilon=
-0.0014^{+0.0014}_{-0.0014}\mbox{ TeV}^{-2}$ and $\epsilon=
-0.0009^{+0.0011}_{-0.0011}\mbox{ TeV}^{-2}$, respectively. Thus,
the signs of the central values in the AA model agree with our
results, whereas the uncertainties are of the same order. From the
carried out analysis it follows that the AA model is mainly
responsible for signals of the Abelian $Z'$ gauge boson although a
lot of details concerning its interactions is not accounted for
within this fit.

The $Z'$ boson mass is related to the contact interaction scale
$\epsilon_0$ by Eq. (\ref{5}). If the $Z'$ boson couples to the SM
particles with a strength comparable with the electroweak forces
$\tilde{g}\simeq g$, the central values of $\bar\epsilon_0$
correspond to the masses of order 3--4 TeV, whereas the lower
limit on $m_{Z'}$ is about 1.5--1.7 TeV. Thus, although the $Z'$
boson is not detected at LEP, it could be light enough to be
discovered at LHC.

\section{Acknowledgement}

A.G. thanks D.Kazakov, S.Mikhailov and O.Teryaev for fruitful
discussions and the JINR for its hospitality while this work was
being completed.

\begin{table}
\centering \caption{95\% confidence level lower limits on the $Z'$
mass for some popular models.}\label{t1}
\begin{tabular}{c|c|c|c|c|c}\hline
 Model & $\chi$ & $\psi$ & $\eta$ & L--R & SSM \\ \hline
 $m^{\rm limit}_{Z'}, {\rm GeV}/c^2$ &
 678 & 463 & 436 & 800 & 1890 \\
 \hline
\end{tabular}\end{table}

\begin{table}\label{t2}\centering
\caption{The boundary angle $z^\ast$ and the observable
$\Delta\sigma_l(z^\ast)$ (pb) computed at energies of the LEP
experiments.}
\begin{tabular}{c|c|r|r}
\hline
 $\sqrt{s}$, GeV & $z^*$ & $\Delta\sigma_\mu(z)$, pb &
 $\Delta\sigma_\tau(z)$, pb  \\ \hline
 130 & 0.486 & $-0.053 \pm 0.398$ & $-0.196 \pm 0.501$ \\
 136 & 0.464 & $ 0.313 \pm 0.364$ & $ 0.352 \pm 0.534$ \\
 161 & 0.406 & $-0.278 \pm 0.262$ & $ 0.373 \pm 0.319$ \\
 172 & 0.391 & $ 0.178 \pm 0.271$ & $-0.834 \pm 0.315$ \\
 183 & 0.379 & $-0.042 \pm 0.107$ & $ 0.079 \pm 0.138$ \\
 189 & 0.374 & $-0.026 \pm 0.060$ & $ 0.023 \pm 0.080$ \\
 192 & 0.372 & $-0.096 \pm 0.142$ & $ 0.065 \pm 0.194$ \\
 196 & 0.369 & $ 0.047 \pm 0.082$ & $-0.174 \pm 0.117$ \\
 200 & 0.366 & $-0.065 \pm 0.078$ & $ 0.003 \pm 0.109$ \\
 202 & 0.364 & $-0.059 \pm 0.118$ & $ 0.067 \pm 0.150$ \\
 205 & 0.362 & $-0.034 \pm 0.086$ & $ 0.034 \pm 0.106$ \\
 207 & 0.361 & $-0.029 \pm 0.068$ & $ 0.003 \pm 0.089$ \\
\hline
\end{tabular}
\end{table}

\begin{table}
\caption{
The fitted values of the contact coupling $\epsilon_0$ and their 68\%
confidence level uncertainties, the 95\% confidence
level lower limit on the scale $\Lambda$, the probability of the $Z'$
signal, $P$, and the fitted values of $\zeta=m^2_Z/m^2_{Z'}$ from the
analysis of terms $\sim m^{-4}_{Z'}$.
}\label{t6}\centering
\begin{tabular}{l|c|c|c|c}
\hline Data set & $\epsilon_0$ & $\Lambda$, TeV & $P$ & $\zeta$ \\
\hline
$\mu\mu$ & $0.0000455^{+0.0000459}_{-0.0000462}$ & 16.3 & 0.83 &
 $0.006\pm 0.213$ \\
$\mu\mu$ and $\tau\tau$ & $0.000030 \pm 0.000037$ & 18.7 & 0.79 &
 $-0.029\pm 0.231$ \\
\hline
\end{tabular}
\end{table}

\end{document}